\documentclass[12pt,a4paper]{article}
\usepackage{amsfonts,amssymb}
\usepackage{amsmath}
\newcommand{\EQ}{\begin{equation}}
\newcommand{\EN}{\end{equation}}

\newtheorem{theo}{Theorem}[section]
\newtheorem{corollary}{Corollary}
\newtheorem{lem}{Lemma}
\newtheorem{prop}{Proposition}

\newtheorem{deff}{Definition}
\newtheorem{rem}{Remark}

\newcommand{\zero}{{\mathbf{0}}}

\newcommand{\supp}{\mbox{\rm supp}}
\newcommand{\wt}{\mbox{\rm wt}}
\newcommand{\w}{\mbox{\rm wt}}

\newcommand{\bc}{{\bf c}}

\newcommand{\by}{{\bf y}}
\newcommand{\bx}{{\bf x}}
\newcommand{\bz}{{\bf z}}
\newcommand{\bv}{{\bf v}}
\newcommand{\bu}{{\bf u}}

\newcommand{\Z}{\mathbb{Z}}
\newcommand{\F}{\mathbb{F}}

\newcommand{\IA}{\operatorname{IA}}

\newcommand{\pr}{\noindent{\bf Proof. \ }}
\newcommand{\qed}{\hspace*{5 mm}$\Box$}

\usepackage[pagewise]{lineno}

\title{On self-dual completely regular codes with covering radius $\rho\leq 3$}

\author{J. Borges and V. A. Zinoviev}

\begin{document}
\maketitle

\centerline{\scshape Joaquim Borges}
\medskip
{\footnotesize
 \centerline{Department of Information and Communications Engineering}
 \centerline{Universitat Aut\`{o}noma de Barcelona}
} 

\medskip

\centerline{\scshape Victor Zinoviev}
\medskip
{\footnotesize
 \centerline{A.A. Kharkevich Institute for Problems of Information Transmission}
 \centerline{Russian Academy of Sciences}
} 

\bigskip

\begin{abstract}
We give a complete classification of self-dual completely regular codes
with covering radius $\rho \leq 3$. For $\rho=1$ the results are almost trivial. For $\rho=2$, by using properties of the more general class of uniformly packed codes in the wide sense, we show that there are two sporadic such codes, of length $8$, and an infinite family, of length $4$, apart from the direct sum of two self-dual completely regular codes with $\rho=1$, each one. For $\rho=3$, in some cases, we use similar techniques to the ones used for $\rho=2$. However, for some other cases we use different methods, namely, the Pless power moments which allow to us to discard several possibilities.
We show that there are only two self-dual completely regular codes with $\rho=3$ and $d\geq 3$, which are both ternary: the extended ternary Golay code and the direct sum of three ternary Hamming codes of length 4. Therefore, any self-dual completely regular  code with $d\geq 3$ and $\rho=3$ is ternary and has length 12.

We provide the intersection arrays for all such codes.
\end{abstract}

\section{Introduction}
Denote by $\F^n_q$ the $n$-dimensional vector space over the finite field of order $q$, where $q$ is a prime power. The {\em (Hamming) distance} between two vectors $\bv,\bu\in\F_q^n$, denoted by $d(\bv,\bu)$, is the number of coordinates in which they differ. The {\em (Hamming) weight} of a vector $\bv\in\F_q^n$, denoted by $\w(\bv)$, is the number of nonzero coordinates of $\bv$.

A $q$-ary {\em code} $C$ of length $n$ is a subset $C\subseteq\F_q^n$. The elements of $C$ are called {\em codewords}. The {\em minimum distance} $d$ of $C$ is the minimum distance between any pair of codewords. The {\em minimum weight} $w$ of $C$ is the minimum weight of any nonzero codeword. A {\em linear code} with parameters $[n,k,d]_q$ is a $q$-ary code of length $n$ with minimum distance $d$, such that it is a $k$-dimensional subspace of $\F_q^n$. For linear codes, the minimum distance and the minimum weight coincide, $d=w$. A $t$-{\em weight} code is a code where the nonzero codewords have $t$ different weights ($t\geq 1$). A linear code of length $n$ is said to be {\em antipodal} if there is some codeword of weight $n$.

The {\em packing radius} of a code $C$ is
 $e=\lfloor (d-1)/2 \rfloor$. Given any vector $\bv \in \F_q^n$, its
distance to the code $C$ is $d(\bv,C)=\min_{\bx \in C}\{
d(\bv, \bx)\}$ and the {\em covering radius} of the code $C$ is
$\rho=\max_{\bv \in \F_q^n} \{d(\bv, C)\}$. Note that $e\leq \rho$. If $e=\rho$, then $C$ is a {\em perfect code}.
 It is well known that any nontrivial (with more than two codewords) perfect code has $e\leq 3$ \cite{Tiet,ZL73}. For $e=1$, linear perfect codes are called Hamming codes which exist for lengths $n=(q^m-1)/(q-1)$ ($m\geq 2$), dimension $k=n-m$ and minimum distance $d=3$. For $e=2$, the only nontrivial perfect code is the ternary Golay code of length 11.

Given two vectors $\bv=(v_1,\ldots,v_n)$ and $\bu=(u_1,\ldots,u_n)$, their inner product is
$$
\bv\cdot\bu=\sum_{i=1}^{n} v_iu_i\in\F_q.
$$
For a linear code $C$, its {\em dual code} is $C^\perp=\{\bx\in\F_q^n \mid \bx\cdot \bv=0,\; \forall \bv\in C\}$. The code $C$ is {\em self-dual} if $C=C^\perp$. In this case, $C$ and $C^\perp$ have the same dimension $n/2$, hence $n$ must be even.

Denote by $\zero$ the all-zero vector. The {\em support} of a vector $\bx=(x_1,\ldots,x_n)\in\F_q^n$ is the set of nonzero coordinate positions of $\bx$, $\supp(\bx)=\{i\in\{1,\ldots,n\}\mid x_i\neq 0\}$. Say that a vector $\bx=(x_1,\ldots,x_n)\in\F_q^n$ {\em covers} a vector $\by=(y_1,\ldots,y_n)\in\F_q^n$ if $x_i=y_i$, for all $i=1,\ldots,n$ such that $y_i\neq 0$.

For a given code $C$ of length $n$ and covering radius $\rho$,
define
\[
C(i)~=~\{\bx \in \F_q^n:\;d(\bx,C)=i\},\;\;i=0,1,\ldots,\rho.
\]
The sets $C(0)=C,C(1),\ldots,C(\rho)$ are called the {\em subconstituents} of $C$.

Say that two vectors $\bx$ and $\by$ are {\em neighbors} if
$d(\bx,\by)=1$.

\begin{deff}[\cite{N92}]\label{de:1.1} A code $C$ of length
$n$ and covering radius $\rho$ is {\em completely regular} (shortly
CR), if for all $l\geq 0$ every vector $\bx \in C(l)$ has the same
number $c_l$ of neighbors in $C(l-1)$ and the same number $b_l$ of
neighbors in $C(l+1)$. Define $a_l = (q-1){\cdot}n-b_l-c_l$ and
set $c_0=b_\rho=0$. The parameters $a_l$, $b_l$ and
$c_l$ ($0\leq l\leq \rho$) are called {\em intersection numbers}
and the sequence $\{b_0, \ldots, b_{\rho-1}; c_1,\ldots, c_{\rho}\}$
is called the {\em intersection array} (shortly $\IA$) of $C$.
\end{deff}

For any $\bv\in\F^n_q$ and any $t\in\{0,\ldots,n\}$, define $B_{\bv,t}=|\{x\in C\mid d(v,x)=t\}|$. Then, $C$ is CR if $B_{\bv,t}$ depends only on $t$ and $d(\bv,C)$ \cite{D73}. The equivalence with Definition \ref{de:1.1} can be seen in \cite{N92}.


%

Existence, construction and classification of completely regular codes, in general, are open hard problems (see \cite{BCN89,DKT,KKM,N92}) of algebraic and combinatorial coding theory.

All linear completely regular codes with covering radius
$\rho=1$ are known \cite{BRZ10}. The next case, i.e.
completely regular codes with $\rho = 2$, was solved for the special
case when the dual codes are antipodal \cite{BRZ10,rho2}. In the present paper, we classify all self-dual completely regular codes with covering radius $\rho\leq 3$.

In Section \ref{Basic}, we see some definitions and results that we use later. In Section \ref{Classif2}, we show that for $\rho=1$ we only have some trivial codes with length and minimum distance $n=d=2$, and the ternary Hamming code of length 4. For $\rho=2$, we prove that the only possible parameters for a self-dual completely regular code are: $[8,4,4]_2$, $[8,4,3]_3$, and $[4,2,3]_q$ for any $q=2^r$ with $r>1$, apart from the direct sum of two self-dual $[2,1,2]_q$ codes. In Section \ref{Classif}, we prove  that for $\rho=3$ the only possibilities are: the direct sum of three self-dual $[2,1,2]_q$ codes, a $[12,6,6]_3$ code and a $[12,6,3]_3$ code. We identify all such codes and show that, indeed, they are self-dual and completely regular. Moreover, all such codes are antipodal except when they are direct sums of other codes. Finally, in Section \ref{concluding}, we summarize the results and briefly discuss the case $\rho > 3$.

\section{Definitions and preliminary results}\label{Basic}

In this section we see several results we will need in the next section.

\subsection{CR and UPWS codes}

A $q$-ary $t-(n,m,\lambda)$-{\em design} is a collection $S$ of vectors of weight $m$ in $\F_q^n$ with the property that every vector
$\bv$ of weight $t$ is covered by exactly $\lambda$ vectors $\by\in S$ ($t\leq m\leq n$). As can be seen in \cite{GvT}, any $q$-ary $t-(n,m,\lambda)$-design is also a $q$-ary $i-(n,m,\lambda_i)$-design for $0\leq i\leq t$, where
\begin{equation}\label{idesign}
  \lambda_i=\lambda\frac{\binom{n-i}{t-i}}{\binom{k-i}{t-i}}(q-1)^{t-i}.
\end{equation}

\begin{lem}[\cite{GvT}]\label{disseny}
  Let $C$ be a CR code with packing radius $e$ and containing the all-zero vector. Then the codewords of any nonzero weight $w$ form a $q$-ary $e$-design and even a $q$-ary $(e+1)$-design if the minimum distance is $d=2e+2$.
\end{lem}

Now, we see an easy but fundamental property. For a code $C$ of length $n$, denote by $C_w$ the set of codewords of weight $w$.

\begin{lem}\label{support}
If $C$ is a CR code of length $n$, containing the zero codeword, and with minimum weight $d$, then $\bigcup_{\bx\in C_d}\supp(\bx)=\{1,\ldots,n\}$.
\end{lem}

\pr
Otherwise taking a 1-weight vector $\bv$, we would have that $B_{\bv,d-1}>0$ if the nonzero coordinate is in $\bigcup_{\bx\in C_d}\supp(\bx)$, but $B_{\bv,d-1}=0$ if not. Hence, $C$ would not be CR.
\qed

\begin{rem}
  Lemma \ref{support} can be also proven taking into account that the codewords in $C_d$ form a $q$-ary $e$-design (see Lemma \ref{disseny}).
\end{rem}

The next property is a well-known construction of CR codes by direct sum. Recall that the direct sum of two codes $C_1$ and $C_2$ is defined as
$$
C_1 \oplus C_2 =\{(\bx,\by)\mid \bx\in C_1, \by\in C_2\}.
$$

If $C_1$ and $C_2$ are linear codes, then $C_1 \oplus C_2$ is a linear code with generator matrix:
$$
G=\left(
    \begin{array}{c|c}
      G_1 & \zero \\
      \hline
      \zero & G_2 \\
    \end{array}
  \right),
$$
where $G_1$ is a generator matrix for $C_1$ and $G_2$ is a generator matrix for $C_2$.

\begin{lem}[\cite{BZZ74}]\label{DirectSum}
  Let $j$ be a positive integer and let $C_i$, $i=1,\ldots,j$ be $q$-ary CR codes with the same length, dimension, minimum distance, with covering radius $\rho=1$ and intersection array $\IA=(b_0,c_1)$. Then, the direct sum $C=C_1\oplus\cdots\oplus C_j$ is a CR code with covering radius $j$ and intersection array
  $$
  \IA=\{jb_0,(j-1)b_0,\ldots,b_0;c_1,2c_1,\ldots,jc_1\}.
  $$
\end{lem}

\begin{deff}[\cite{BZZ74}]\label{DefUPWS}
A code $C\subseteq\F_q^n$ with covering radius $\rho$ is {\em uniformly packed in the wide sense (UPWS)} if there exist rational numbers $\beta_0,\ldots,\beta_\rho$ such that
\begin{equation}\label{SumUPWS}
\sum_{i=0}^{\rho} \beta_i B_{\bx,i} = 1,
\end{equation}
for any $\bx\in\F^n_q$. The numbers $\beta_0,\ldots,\beta_\rho$ are called the {\em packing coefficients}.
\end{deff}

For UPWS codes, there is a generalized version of the celebrate sphere packing condition for perfect codes.

\begin{lem}[\cite{BZZ74}]\label{SPC}
Let $C\subseteq\F^n_q$ be a UPWS code with covering radius $\rho$ and packing coefficients $\beta_0,\ldots,\beta_\rho$. Then
\begin{equation*}
  |C|=\frac{q^n}{\sum_{i=0}^{\rho}\beta_i(q-1)^i\binom{n}{i}}.
\end{equation*}
\end{lem}

For a linear code $C$, denote by $s$ the number of nonzero weights of $C^\perp$. Following to Delsarte \cite{D73}, we call {\em external distance} the parameter $s$.

\begin{lem}\label{props}
Let $C$ be a linear code with covering radius $\rho$, packing radius $e$ and external distance $s$.
\begin{itemize}
  \item[(i)] $\rho\leq s$ \cite{D73}.
  \item[(ii)] $\rho = s$ if and only if $C$ is UPWS \cite{BZZ77}.
  \item[(iii)] If $C$ is CR, then $\rho=s$ \cite{S90}.
  \item[(iv)] If $C$ is UPWS and $\rho=e+1$, then $C$ is CR \cite{GvT,SZZ71}.
\end{itemize}
\end{lem}

Let $C$ be a CR code. Set $p_{i,j}=B_{\bv,j}$, for any $\bv$ such that $d(\bv,C)=i$ ($0\leq i\leq \rho$).
By Lemma \ref{props}, any CR code is also a UPWS code. Hence, for any CR code we can apply Lemma \ref{SPC}.
\begin{prop}\label{PP}
Let $C$ be a CR $[n,k,d]_q$ code with covering radius $\rho>1$. Then, the packing coefficients verify:
\begin{itemize}
  \item[(i)] If $d=2\rho$, then
  \begin{equation}\label{2r}
\beta_0=\ldots =\beta_{\rho-1}=1;\;\;\;\;\;\;\;\;\beta_\rho=\frac{q^{n-k}-\sum_{i=0}^{\rho-1}(q-1)^i \binom{n}{i}}{(q-1)^\rho\binom{n}{\rho}}.
  \end{equation}
  \item[(ii)] If $d=2\rho-1$, then
  \begin{equation}\label{2r-1}
\beta_0=\ldots =\beta_{\rho-2}=1;\;\beta_{\rho-1}+\beta_\rho p_{\rho-1,\rho}=1;\;\beta_\rho=\frac{q^{n-k}-\sum_{i=0}^{\rho-1}(q-1)^i \binom{n}{i}}{(q-1)^\rho\binom{n}{\rho}-p_{\rho-1,\rho}(q-1)^{\rho-1}\binom{n}{\rho-1}}.
  \end{equation}
  \item[(iii)] If $d=2\rho-2$, then
\begin{eqnarray}\nonumber
  \beta_0 &=& \ldots=\beta_{\rho-3}=1;\;\beta_{\rho-2}+\beta_\rho p_{\rho-2,\rho}=1;\;\beta_{\rho-1}p_{\rho-1,\rho-1}+\beta_\rho p_{\rho-1,\rho}=1; \\ \label{2r-2}
  \beta_\rho &=& \frac{q^{n-k}-\sum_{i=0}^{\rho-2}(q-1)^i\binom{n}{i} - p_{\rho-1,\rho-1}^{-1} \binom{n}{\rho-1}(q-1)^{\rho-1}}{ \binom{n}{\rho}(q-1)^\rho-p_{\rho-1,\rho}p_{\rho-1,\rho-1}^{-1}\binom{n}{\rho-1}(q-1)^{\rho-1}-p_{\rho-2,\rho}(q-1)^{\rho-2}\binom{n}{\rho-2}}.
\end{eqnarray}
\end{itemize}

In all cases $\beta_\rho^{-1}$ is a natural number.
\end{prop}

\pr
(i) Since $C$ is CR, $C$ is also UPWS. For any $i=0,\ldots,\rho-1$, we have that $p_{i,i}=1$ because $i < \rho=d/2$. Moreover, for any $j\in\{0,\ldots,\rho\}\setminus\{i\}$, $p_{i,j}=0$. Indeed, if $d(\bx,C)=i$ and $\bc,\bc'\in C$ are such that $d(\bc,\bx)=i$ and $d(\bc',\bx)=j$, then $d(\bc,\bc')\leq i+j < d$ which is a contradiction. Hence, according to Eq. (\ref{SumUPWS}) in Definition \ref{DefUPWS}, we have $\beta_i=1$ for each $i=0,\ldots,\rho-1$.

Therefore, by Lemma \ref{SPC}, it follows that
$$
|C|=q^k=\frac{q^n}{\sum_{i=0}^{\rho-1}(q-1)^i\binom{n}{i}+\beta_\rho (q-1)^\rho \binom{n}{\rho}}\;\Longrightarrow\;\beta_\rho=\frac{q^{n-k}-\sum_{i=0}^{\rho-1}(q-1)^i \binom{n}{i}}{(q-1)^\rho\binom{n}{\rho}}.
$$

(ii) Now, for any $i=0,\ldots,\rho-1$, we have again that $p_{i,i}=1$ because $i \leq \rho-1<d/2$. Moreover, for any $j\in\{0,\ldots,\rho-1\}\setminus\{i\}$, $p_{i,j}=0$. Indeed, if $d(\bx,C)=i$ and $\bc,\bc'\in C$ are such that $d(\bc,\bx)=i$ and $d(\bc',\bx)=j$, then $d(\bc,\bc')\leq i+j < d$ which is a contradiction. Hence, according to Eq. (\ref{SumUPWS}) in Definition \ref{DefUPWS}, we have $\beta_i=1$ for each $i=0,\ldots,\rho-2$. Thus, $p_{\rho-1,\rho-1}=1$ and $\beta_{\rho-1}+\beta_\rho p_{\rho-1,\rho}=1$.

Again by Lemma \ref{SPC} and using $\beta_{\rho-1}=1-\beta_\rho p_{\rho-1,\rho}$, Eq. (\ref{2r-1}) is obtained.

(iii) In this case, and by similar arguments, we have $p_{i,i}=1$ for any $i=0,\ldots,\rho-2$. For any $j\in\{0,\ldots,\rho-1\}\setminus\{i\}$, $p_{i,j}=0$. Indeed, if $d(\bx,C)=i$ and $\bc,\bc'\in C$ are such that $d(\bc,\bx)=i$ and $d(\bc',\bx)=j$, then $d(\bc,\bc')\leq i+j < d$ which is a contradiction. Hence, we have $\beta_i=1$ for all $i=0,\ldots,\rho-3$. Thus, $p_{\rho-2,\rho-2}=1$ and $\beta_{\rho-2}+\beta_\rho p_{\rho-2,\rho}=1$, since $p_{\rho-2,\rho-1}=0$. On the other hand, $\beta_{\rho-1}p_{\rho-1,\rho-1}+\beta_\rho p_{\rho-1,\rho}=1$.

Again by Lemma \ref{SPC}, using $\beta_{\rho-2}=1-\beta_\rho p_{\rho-2,\rho}$ and  $\beta_{\rho-1}=1-(\beta_\rho p_{\rho-1,\rho})/p_{\rho-1,\rho-1}$, Eq. (\ref{2r-2}) is obtained.

In every case (i), (ii) and (iii), it is clear that $p_{\rho,i}=0$, for all $i=0,\ldots,\rho-1$ and thus $\beta_\rho p_{\rho,\rho}=1$ and $\beta_\rho^{-1}$ is a natural number.
\qed

We are interested in the case when $C$ is self-dual and CR with covering radius $\rho=2$ or $\rho=3$.

\begin{corollary}\label{PP2}
  Let $C$ be a self-dual CR $[2k,k,4]_q$ code with covering radius $\rho=2$. Then, the packing coefficient $\beta_2$ is:
$$
\beta_2=\frac{q^k-1-2k(q-1)}{(q-1)^2 k(2k-1)},
$$
and $\beta_2^{-1}$ is a natural number.
\end{corollary}

\pr
Straightforward substituting $\rho=2$ and $n=2k$ in Eq. (\ref{2r}).
\qed

\begin{corollary}\label{PP3}
  Let $C$ be a self-dual CR $[2k,k,d]_q$ code with covering radius $\rho=3$. Then, the packing coefficient $\beta_3$ is:
\begin{itemize}
  \item[(i)] If $d=6$, then
  $$
  \beta_3=3\frac{q^k-1-2k(q-1)-k(2k-1)(q-1)^2}{(q-1)^3k(2k-1)(2k-2)}.
  $$
  \item[(ii)] If $d=5$, then
  $$
  \beta_3=3\frac{q^k-1-2k(q-1)-k(2k-1)(q-1)^2}{k(2k-1)(q-1)^2[(2k-2)(q-1)-p_{2,3}]},
  $$
  where $0\leq p_{2,3} \leq \frac{2(q-1)(k-1)}{3}$.
  \item[(iii)] If $d=4$, then
  $$
  \beta_3=3\frac{(\lambda+1)(q^k-1)-k(q-1)[2(\lambda+1)+(2k-1)(q-1)]}
  {k(2k-1)(q-1)^2[(\lambda+1)(2k-2)(q-1)-2\lambda(\lambda+1)-6\lambda(q-2)-3\lambda']},
  $$
  where $\lambda=p_{2,2}-1$ and $\lambda'=p_{2,3}-2\lambda (q-2)$. Moreover, $1\leq\lambda\leq k-1$.
\end{itemize}
\end{corollary}

\pr
(i) Put $\rho=3$ and $n=2k$ in Eq. (\ref{2r}).

(ii) Again Put $\rho=3$ and $n=2k$ in Eq. (\ref{2r-1}).

A CR $[n,k,d]_q$ code with $\rho=e+1$ is a quasi-perfect uniformly packed code \cite{GvT}.
In this case, as can be seen in \cite{BZZ74}, the packing parameters verify:
$$
\beta_0=\beta_1=1;\;\;\ldots\;\;\beta_{e}=1-s/m;\;\;\beta_{e+1}=1/m;
$$
where $m$ and $s$ are integer values in the ranges:
$$
0< m < \frac{n(q-1)}{e+1}\;\;\;\;\;\;\;\;  0\leq s \leq \frac{(q-1)(n-e)}{e+1}.
$$

Since $d=5$ and $\rho=3$, we are in the case of a quasi-perfect uniformly packed code.
Therefore, $s=p_{2,3}$ and $m=p_{3,3}$. Then, it follows the bound for $p_{2,3}$.

(iii) By Lemma \ref{disseny}, the codewords in $C_4$ form a $q$-ary $2-(2k,4,\lambda)$-design. Consider a $2$-weight vector $\bv$. Such vector is covered by $\lambda$ codewords in $C_4$ and it is also at distance $2$ from de zero codeword. Thus, $p_{2,2}=\lambda+1$. The codewords at distance $3$ from $\bv$ are:
\begin{itemize}
  \item[(a)] The $\mu$ codewords in $C_4$ containing the support of $\bv$ and covering just one of the nonzero coordinates of $\bv$.
  \item[(b)] The $\lambda'$ codewords in $C_5$ (if $C_5$ is not empty) covering $\bv$.
\end{itemize}
For (a), let $X=\{\bx\in C_4\mid \supp(\bv)\subset \supp(\bx)\}$. There are $(q-1)^2$ vectors with the same support that $\bv$. Each one of these vectors is covered by $\lambda$ vectors in $C_4$. Thus, $|X|=\lambda (q-1)^2$. Let $Y\subset C_4$ be the set of codewords of weight $4$ which are multiples of some codeword in $C_4$ covering $\bv$. Clearly, $|Y|=\lambda (q-1)$. Hence, for any codeword in $X\setminus Y$, we have two multiples that cover exactly one nonzero coordinate of $\bv$. This means that
$$
\mu=2\frac{|X|\setminus |Y|}{q-1}=2[\lambda (q-1)-\lambda]=2\lambda (q-2).
$$

For (b), simply consider that $C_5$ form a $2-(2k,5,\lambda')$-design. If $C_5=\emptyset$, then we set $\lambda'=0$.
As a consequence, we have that $p_{2,3}=2\lambda (q-2) + \lambda'$.
Now consider a $1$-weight vector $\bu$. The codewords at distance $3$ from $\bu$ are those in $C_4$ covering $\bu$. According to Eq. (\ref{idesign}), such number of vectors is $p_{1,3}=\lambda (2k-1)(q-1)/3$.

Substituting $p_{1,3}=\lambda (2k-1)(q-1)/3$, $p_{2,2}=\lambda+1$, and $p_{2,3}=2\lambda (q-2) + \lambda'$ in Eq. (\ref{2r-2}), we obtain the expression for $\beta_3$.

Finally, note that if $\bx,\by\in C_4$ are codewords covering the $2$-weight vector $\bv$, then $\supp(\bx)\cap \supp(\by)=\supp(\bv)$ (otherwise $\wt(\bx-\by)<4=d$). The union of the supports of the $\lambda$ vectors covering $\bv$ must have cardinality at most $n=2k$. Therefore, $2+2\lambda\leq 2k$, implying $\lambda\leq k-1$.
\qed

\subsection{Self-dual two-weight and three-weight  codes}

We start with three general an easy results on self-dual codes.

\begin{lem}\label{q23}
Let $C$ be a $q$-ary self-dual code.
\begin{itemize}
\item[(i)] If $q=2$, then the weight of any codeword is even.
\item[(ii)] If $q=3$, then the weight of any codeword is divisible by $3$.
\end{itemize}
\end{lem}

\pr
If $C$ is self-dual, then $\bx\cdot \bx=0$, for any codeword $\bx\in C$. Therefore (i) is trivial. For (ii), note that for any ternary vector $\bz\in\F_3^n$, $\bz\cdot \bz \equiv \w(\bz) \pmod 3$.
\qed

The next well-known property shows which is the only self-dual perfect code.

\begin{lem}\label{HammingSelf}
The only self-dual perfect code is the ternary Hamming $[4,2,3]_3$ code.
\end{lem}

\pr
For any self-dual $[n,k,d]_q$ code, we have that $n=2k$.
The only perfect codes with minimum distance $d>3$ are the ternary Golay $[11,6,5]_3$ code, the binary Golay $[23,12,7]_2$ code and binary repetition codes $[n,1,n]_2$ codes of odd length. Since the length of these codes is odd, no one can be self-dual.

For the case of a self-dual perfect code with $d=3$, hence for a self-dual Hamming $[n,n-m,3]_q$ code, $n=2(n-m)$ and thus $n=2m$ implying
$$
\frac{q^m-1}{q-1}=2m.
$$
The only solution is $q=3$ and $m=2$. Therefore, $n=4$ and $k=2$.
\qed

\begin{lem}\label{intersect1}
  If $C$ is a self-dual code, then $|\supp(\bx)\cap\supp(\by)|\neq 1$, for any $\bx,\by\in C$.
\end{lem}

\pr
Otherwise $\bx$ and $\by$ would not be orthogonal vectors.
\qed

Now, we show the nonexistence of a particular self-dual code.

\begin{lem}\label{NoBush}
There is no self-dual $[6,3,4]_4$ code.
\end{lem}

\pr
Let $C$ be a $[6,3,4]_4$ code and consider a generator matrix for $C$:
$$
G=\left(
    \begin{array}{c|c}
      I_3 & P \\
    \end{array}
  \right),
$$
where $I_3$ is the $3\times 3$ identity matrix and $P$ is a $3\times 3$ matrix with nonzero entries, since $C$ has minimum weight $4$.
If $C$ is self-dual then any row of $G$ must be self-orthogonal, implying that for any row $abc$ of $P$, we have $1+a^2+b^2+c^2=0$. Thus, $(a+b+c)^2=1$ and $a+b+c=1$. In $\F_4$ and since $P$ has no zero entries, this means that $abc\in\{1xx,x1x,xx1\}$, where $x\neq 0$. If $abc=111$, then it is not orthogonal to any other row
$a'b'c'\in\{1xx,x1x,xx1\}$, where $x\notin\{0,1\}$. So, the three rows contain exactly one $1$. Hence, two rows of $P$ have the same value for $x$, say $\alpha$. But such two rows must agree in one position (with $1$), since the distance must be two. Therefore, they are not orthogonal and hence the corresponding rows of $G$ are also non-orthogonal.
\qed

 For a code $C$, let $A_w=|C_w|$. Thus, $\{A_0,A_1,\ldots,A_n\}$ is the weight distribution of $C$. The Pless power moments \cite{Pless}, as well as the McWilliams identities, relate the weight distribution of $C$ and the weight distribution of $C^\perp$, for a linear code $C$. The first five Pless power moments can be seen in \cite[pp. 259-260]{HP}. For a self-dual 3-weight $[2k,k,d]_q$ code with $d\geq 3$ and nonzero weights $w_1$, $w_2$ and $w_3$ the first three equations are:

 \begin{eqnarray}\label{PPM1}
A_{w_1}+A_{w_2}+A_{w_3} &=& q^k-1 \\ \label{PPM2}
w_1A_{w_1}+w_2A_{w_2}+w_3A_{w_3} &=& q^{k-1}2k(q-1) \\ \label{PPM3}
w_1^2A_{w_1}+w_2^2A_{w_2}+w_3^2A_{w_3} &=& q^{k-2}\left[2k(q-1)(2k(q-1) +1)\right].
\end{eqnarray}

As a consequence of these equations, we have the following result.

\begin{lem}\label{k3}
If $C$ is a self-dual $3$-weight $[2k,k,d]_q$ code with nonzero weights $w_1,w_2,w_3$ such that $3\leq d=w_1 < w_2 < w_3$, then $q(2k-w_3)<2k$.
\end{lem}

\pr
Combining Eqs. (\ref{PPM1}) and (\ref{PPM2}), we get
$$
(w_3-w_1)A_{w_1} + (w_3-w_2)A_{w_2} = w_3(q^k-1)- q^{k-1}2k(q-1),
$$
which gives
$$
(w_3-w_1)A_{w_1} + (w_3-w_2)A_{w_2} + w_3 = q^{k-1}[(w_3-2k)q + 2k].
$$
Obviously, both hand sides must be positive. Thus, we obtain $q(2k-w_3)<2k$.
\qed

\begin{rem}
Lemma \ref{k3} can be easily generalized for any self-dual code with $d\geq 3$. With the same argument, one obtains $q(2k-w_r)<2k$, where $w_r$ is the greatest nonzero weight.
\end{rem}

Note that for any $q$-ary linear code, $A_w$ is a multiple of $q-1$ (indeed, given any codeword, its $q-1$ multiples are codewords). Hence, we define $B_w=A_w/(q-1)$. Therefore, after dividing each term by $q-1$, Eqs. (\ref{PPM1}), (\ref{PPM2}) and (\ref{PPM3}) become:
\begin{eqnarray}\label{PPMB1}
B_{w_1}+B_{w_2}+B_{w_3} &=& \frac{q^k-1}{q-1} \\ \label{PPMB2}
w_1B_{w_1}+w_2B_{w_2}+w_3B_{w_3} &=& q^{k-1}2k \\ \label{PPMB3}
w_1^2B_{w_1}+w_2^2B_{w_2}+w_3^2B_{w_3} &=& q^{k-2}2k(2k(q-1) +1).
\end{eqnarray}

We shall solve the system of Eqs. (\ref{PPMB1}), (\ref{PPMB2}) and (\ref{PPMB3}) for several different cases. Therefore, we summarize in  Table \ref{taula} some results we need.

\begin{table}[h!]
\centering
  \begin{tabular}{|c|c|c|c|c|c|}
  \hline
  $w_1$ & $w_2$  & $w_3$ & $q$ & $k$ & $(B_{w_1},B_{w_2},B_{w_3})$ \\
  \hline
  5 & 6 & 7 & 7 & 4 & $(168,-280,512)$ \\
  5 & 6 & 8 & 7 & 4 & $(-8/3,232,512/3)$ \\
  5 & 7 & 8 & 7 & 4 & $(224/3,232,280/3)$ \\
  3 & 4 & 5 & $q$ & 3 & $\left(q^2-5q+10,3(-q^2+5q-5),3(q-1)(q-2)\right)$ \\
  4 & 5 & 6 & $q$ & 3 & $\left(15,6(q-4),q^2-5q+10\right)$ \\
  3 & 4 & 6 & $q$ & 3 & $\left(-2(q-4),3(2q-3),(q-1)(q-2)\right)$ \\
  3 & 5 & 6 & $q$ & 3 & $\left(5,3(2q-3),q^2-5q+5\right)$ \\
  \hline
  \end{tabular}
  \caption{Some results of the system of Eqs. (\ref{PPMB1}), (\ref{PPMB2}) and (\ref{PPMB3})}\label{taula}
\end{table}

Directly, from Table \ref{taula}, we can state the nonexistence of certain self-dual 3-weight codes.

\begin{prop}\label{noexis}
The following self-dual $3$-weight codes do not exist:
\begin{itemize}
\item[(i)] A $[8,4,5]_7$ code.
\item[(ii)] A code with nonzero weights $3,4,5$.
\end{itemize}
\end{prop}

\pr
Of course, solving the system of Eqs. (\ref{PPMB1}), (\ref{PPMB2}) and (\ref{PPMB3}) we should obtain positive integer values for $B_1$, $B_2$ and $B_3$.

(i) Let $w_1$, $w_2$ and $w_3$ be the nonzero weights such that $5=w_1 < w_2 < w_3 \leq 8$, then $(w_1,w_2,w_3)\in\{(5,6,7),(5,6,8),(5,7,8)\}$. These cases correspond to the first three rows in Table \ref{taula}. In any case we always have negative and/or noninteger values. Consequently, no one of these codes can exist.

(ii) By Lemma \ref{q23}, $q\geq 4$ and thus, by Lemma \ref{k3}, $k=3$. Then, by Eqs. (\ref{PPMB1}), (\ref{PPMB2}) and (\ref{PPMB3}) we have $B_4=3(-q^2+5q-5)$ (see the fourth row in Table \ref{taula}). Hence, $-q^2+5q-5 > 0$ and thus $q < 4$, which is a contradiction by Lemma \ref{q23}.
\qed

\begin{rem}
  For the case (i) in Proposition \ref{noexis}, note that a $[8,4,5]_q$ code meets the singleton bound ($d\leq n-k+1$) and thus it is a maximum distance separable (MDS) code. The weight distribution of such codes is completely determined and, as can be seen in \cite[p. 320]{MS},
$$
A_d=(q-1)\binom{n}{d}.
$$
For $q=7,n=8,d=5$, this gives $A_5=336$ and hence $B_5=56$, which does not coincide with the results of the system of equations. Therefore, we obtain a contradiction again.
\end{rem}

\section{Self-dual completely regular codes with $\rho\leq 2$}\label{Classif2}

Let $C$ be a self-dual CR $[n,k,d]_q$ code with covering radius $1\leq\rho\leq 2$.
In this section we give a full classification of such codes. Note that $n=2k$ (since $C$ is self-dual) and $C$ is a 1-weight code (or equidistant code) or a 2-weight code (because $s=\rho$ by Lemma \ref{props}). Since $e\leq \rho$, we have that $1\leq d \leq 6$. But for $d\geq 5$, $e=\rho$ and $C$ would be a perfect 2-error-correcting code, that is, $C$ would be a ternary Golay $[11,6,5]_3$ code which obviously is not self-dual (the extended ternary Golay code is self-dual, but with covering radius 3). Clearly, for $d=1$ there is no self-dual code. Therefore, $C$ must be a $[2k,k,d]_q$ code with one weight $d=2$ or with two weights $w_1=d\in\{2,3,4\}$ and $w_2$, where $d < w_2 \leq n$.

Now we study separately the cases $d=2$, $d=3$ and $d=4$.

\subsection{The case $d=2$}

This is a very simple case. If $C$ is a self-dual CR $[n,k,2]_q$ code, then by Lemmas \ref{support} and \ref{intersect1}, $C$ is the direct sum of codes of length 2. If $C_i$ is one such code, then $C_i$ has generator matrix $G_i=(1\;\;\alpha)$, where $1+\alpha^2=0$. Indeed, such a code is CR with covering radius 1 (and $p_{1,1}=2$). Therefore we have the following characterization.

\begin{prop}\label{D2}
  If $C$ is a self-dual CR $[n,k,2]_q$ code, then $C$ is a direct sum $C=C_1\oplus\cdots\oplus C_j$, where $C_i$ is a $[2,1,2]_q$ code ($1\leq i \leq j$) and $q$ is such that $-1$ is a square in $\F_q$. The covering radius of $C$ is $\rho=j$ and its intersection array is:
  $$
  \IA=\{2j(q-1),2(j-1)(q-1),\ldots,2(q-1);2,4,\ldots,2j\}.
  $$
\end{prop}

\pr
Straightforward from Lemma \ref{DirectSum}, taking into account that each $C_i$ has covering radius 1 and intersection array $\{2(q-1);2\}$.
\qed

\subsection{The case $d=3$}

Recall that for any code $C$, the set of codewords of weight $w$ is denoted by $C_w$.

\begin{lem}\label{intersect}
If $C$ is a $2$-weight $[n,k,3]_q$ code such that $|\supp(\bx)\cap\supp(\by)|\in \{0,3\}$ for all $\bx,\by\in C_3$, then $C$ is not CR.
\end{lem}

\pr
Let $\bx,\by\in C_3$ such that $|\supp(\bx)\cap\supp(\by)|= 0$, by the assumption and Lemma \ref{support} such vectors must exist. Then, $C$ has weights 3 and 6. Any other codeword $\bz\in C_3$ will have $\supp(\bz)=\supp(\bx)$ or $\supp(\bz)=\supp(\by)$, otherwise $C$ would have more than two weights.

Without loss of generality, assume that $\bx=(x_1,x_2,x_3,0,0,0)$ and $\by=(0,0,0,y_1,y_2,y_3)$. Now, the vector $\bv=(x_1,v_2,0,0,0,0)$, where $v_2\neq x_2$ is clearly at distance 2 to $C$ and, since $d(\bv,\bx)=d(\bv,\zero)=2$, we have $B_{\bv,2}=2$. Indeed, a vector $\bx'=(x_1,v_2,x_3',0,0,0)$ cannot be a codeword because $d(\bx,\bx')\leq 2$. Now take $\bu=(x_1,0,0,y_1,0,0)$. Clearly, $d(\bu,C)=2$ but $B_{\bu,2}=1$. Therefore, $C$ is not CR.
\qed

\begin{prop}\label{4or8}
If $C$ is a self-dual CR $[n,k,3]_q$ code with covering radius $\rho=2$, then $n=4$ or $n=8$.
\end{prop}

\pr
By Lemmas \ref{support} and \ref{intersect}, there exist codewords $\bx,\by\in C_3$, such that $|\supp(\bx)\cap\supp(\by)|= 2$ and thus $|\supp(\bx)\cup\supp(\by)|= 4$. Now, if $\bz\in C_3$ has $\supp(\bz)\cap(\supp(\bx)\cup\supp(\by))\neq\emptyset$, we claim that $\supp(\bz)\subset(\supp(\bx)\cup\supp(\by))$. Otherwise, without loss of generality assume that $\bx=(1,x_2,x_3,0,\ldots,0)$ and $\by=(1,y_2,0,y_3,0,\ldots,0)$. By (i) of Lemma \ref{intersect}, we can assume that $\bz=(z_1,z_2,0,0,z_3,0,\ldots,0)$. Now, since $q>2$, by Lemma \ref{q23}, we can take a multiple of $\bz$, say $\bz'=(z'_1,z'_2,0,0,z'_3,0,\ldots,0)$, such that $z'_2=x_2-y_2$. Hence, $\wt(\bx-\by-\bz')=4$. But we can take another multiple, say $\bz''$, such that $z''_2\neq x_2-y_2$. In this case, $\wt(\bx-\by-\bz'')=5$. So, $C$ has more than two nonzero weights, leading to a contradiction.

As a consequence, we have that $C_3$ induces a partition of the set of coordinates in $4$-subsets, implying that $n$ is a multiple of $4$. But for $n>8$, clearly $C$ would have more than two nonzero weights. Therefore $n=4$ or $n=8$.
\qed

\subsection{The case $d=4$}

\begin{prop}\label{Case4}
If $C$ is a self-dual CR $[2k,k,4]_q$ code with covering radius $\rho=2$, then $k=4$ and $q=2$.
\end{prop}

\pr
Define the function
$$
f(q,k)=\frac{(q-1)^2 k(2k-1)}{q^k-1-2k(q-1)}.
$$
By Corollary \ref{PP2}, $f(q,k)=\beta_2^{-1}$ and must be a natural number. For $n=2k=4$, $C$ cannot be a 2-weight code with minimum weight 4. Thus $k\geq 3$. Clearly, for a fixed $k$, $f(q,k)$ is a decreasing function on $q$. In fact, it is easy to check that $q$ must be less than 16, otherwise $f(q,k)<1$. Also, fixing $q$, $f(q,k)$ is a decreasing function on $k$ and one obtains that $k\leq 6$. Again, $f(q,k)<1$ for $k>6$. For all these possible values ($2\leq q<16$, $3\leq k\leq 6$), we have computationally checked that the only natural values of $f(q,k)$ are $f(2,4)=4$, $f(2,3)=15$ and $f(4,3)=3$. But $f(2,3)=15$ implies $B_{\bx,2}=15$ which is not possible for $q=2$ and $n=2k=6$. Indeed, if $\bx$ has weight 2, the number of codewords of weight 4 at distance 2 from $\bx$ cannot be greater than 2 and, taking into account de zero codeword we have $B_{\bx,2} \leq 3$.

As a consequence, the only possible values for $q$ and $k$ are $(q,k)\in\{(2,4),(4,3)\}$, but by Lemma \ref{NoBush}, the case $(q,k)=(4,3)$ is not possible.
\qed

\section{Self-dual completely regular codes with $\rho=3$}\label{Classif}

Let $C$ be a self-dual CR $[n,k,d]_q$ code with covering radius $\rho=3$.
In this section we give a full classification of such codes. Note that $n=2k$ (since $C$ is self-dual) and $C$ is a 3-weight code (because $s=\rho=3$ by Lemma \ref{props}). Since $e\leq \rho=3$, we have that $d \leq 8$. But for $d\geq 7$, $e=\rho$ and $C$ would be a perfect 3-error-correcting code, that is, $C$ would be the binary Golay $[23,12,7]_2$ code which obviously is not self-dual (the extended binary Golay code is self-dual, but with covering radius 4). Hence, $C$ must be a $[2k,k,d]_q$ code with weights $w_1=d\in\{2,3,4,5,6\}$, $w_2$ and $w_3$, where $d < w_2 < w_3\leq n$. But the case $d=2$ is trivial: the only possibility is the direct sum of three self-dual $[2,1,2]_q$ codes. Therefore, we study the cases $d\in\{3,4,5,6\}$.

\subsection{The case $d=6$}

\begin{prop}\label{Case6}
If $C$ is a self-dual CR $[2k,k,6]_q$ code with covering radius $\rho=3$, then $k=6$ and $q=3$.
\end{prop}

\pr
Define the function
$$
f(q,k)=\frac{(q-1)^3 k(2k-1)(2k-2)}{3[q^k-1-2k(q-1)-k(2k-1)(q-1)^2]}.
$$
By Corollary \ref{PP3}(i), $f(q,k)=\beta_3^{-1}$ and must be a natural number. For length $2k=n < 8$, $C$ cannot be a 3-weight code with minimum weight 6. Thus $k\geq 4$. Clearly, for a fixed $k$, $f(q,k)$ is a decreasing function on $q$. In fact, it is easy to check that for $q> 53$, $f(q,k)<1$. Also, fixing $q$, $f(q,k)$ is a decreasing function on $k$ and one obtains that $k\leq 10$. Again, $f(q,k)<1$ for $k>10$. For all these possible values ($2\leq q\leq 53$, $4\leq k\leq 10$), we have computationally checked that the only natural values of $f(q,k)$ are $f(7,4)=9$ and $f(3,6)=4$. But $f(7,4)=9$ implies $\beta_3^{-1}=p_{3,3}=9$ which is not possible for $n=2k=8$. Indeed, if $\bx$ has weight 3, the number of codewords of weight 6 at distance 3 from $\bx$ cannot be greater than 1 and, taking into account de zero codeword, we have $p_{3,3} \leq 2$.

As a consequence, the only possible values for $q$ and $k$ are $(q,k)=(3,6)$.
\qed

\subsection{The case $d=5$}

\begin{prop}\label{Case5}
If $C$ is a self-dual CR $[2k,k,5]_q$ code with covering radius $\rho=3$, then $q=7$, $k=4$ and the nonzero weights of $C$ are $w_1=5$, $w_2\in\{6,7\}$ and $w_3=8$.
\end{prop}

\pr
Assume that $C$ is a self-dual CR $[2k,k,5]_q$ code. Since $d=5$, we have $e=2$ and thus $\rho=e+1$. That is, $C$ is a quasi-perfect uniformly packed code \cite{GvT}.

Define the function
$$
g(q,k,s)=\frac{k(2k-1)(q-1)^2[(2k-2)(q-1)-s]}{3[q^k-1-2k(q-1)-k(2k-1)(q-1)^2]}.
$$
Note that $g(q,k,0)$ is the same that $f(q,k)$ in the proof of Proposition \ref{Case6}.
By Corollary \ref{PP3}(ii), putting $s=p_{2,3}$, we have that $g(q,k,s)=\beta_3^{-1}$ and must be a natural number. For length $2k=n < 8$, $C$ cannot be a 3-weight code with minimum weight 5. Thus $k\geq 4$. Clearly, $g(q,k,s)$ is maximum when $s=0$. In this case, as in the proof of Proposition \ref{Case6}, we have that for $q> 53$, $g(q,k,s)<1$. Also, for $k>10$, $g(q,k,s)<1$. For all these possible values ($2\leq q\leq 53$, $4\leq k\leq 10$, and $0\leq s \leq \frac{(q-1)(2k-2)}{3}$, according to Corollary \ref{PP3}(ii)), we have computationally checked that the only natural values of $g(q,k,s)$ are
$$
g(7,4,0)=9;\;\;g(7,4,4)=8;\;\;g(7,4,8)=7;\;\;g(7,4,12)=6.
$$
Hence, in all cases we have $n=2k=8$ and $q=7$.

Now, by Lemma \ref{k3} we obtain $w_3\geq 8$ and since $n=8$, we conclude $w_3=8$.
\qed

\begin{corollary}\label{Nod5}
  There is no self-dual CR $[2k,k,5]_q$ code with covering radius $\rho=3$.
\end{corollary}

\pr
By Proposition \ref{Case5}, such a code would be a $[8,4,5]_7$ code with nonzero weights $(w_1,w_2,w_3)=(5,6,8)$ or $(w_1,w_2,w_3)=(5,7,8)$. The result then follows from Proposition \ref{noexis}.
\qed

\subsection{The case $d=4$}

We start with a very restrictive condition.

\begin{lem}\label{lambda1}
  Let $C$ be a self-dual CR $[2k,k,4]_q$ code with covering radius $\rho=3$. If $q>2$, then $p_{2,2}=2$.
\end{lem}
\pr
Let $\bv=(1,\alpha,0,\ldots,0)$ be a $2$-weight vector. Clearly, $d(\bv,C)=2$ and $\zero$ is a codeword at distance $2$ from $\bv$. assume that $p_{2,2}>2$ and let $\bx,\by\in C_4$ be codewords such that $d(\bv,\bx)=d(\bv,\by)=2$. Then, $\bx$ and $\by$ cover $\bv$. It holds that $\supp(\bx)\cap\supp(\by)=\supp(\bv)$ (otherwise $\wt(\bx-\by)< 4$). Since $\bx$ and $\by$ must be orthogonal vectors, we have that $\alpha^2=-1$. But this should be true for any nonzero element $\alpha\in\F_q^*=\F_q\setminus \{0\}$. However, this only happens in the binary field. Thus, $q=2$.
\qed

Now we establish the nonexistence of self-dual CR quaternary codes of length $n\geq 6$ and minimum distance $d=4$.

\begin{prop}\label{NoQuaternary}
  For $k\geq 3$, there is no self-dual CR $[2k,k,4]_4$ code.
\end{prop}
\pr
If $C$ is a self-dual CR $[2k,k,4]_4$ code, then $C_4$ is a quaternary $2-(2k,4,\lambda)$-design, by Lemma \ref{disseny} (where $\lambda=1$, by Lemma \ref{lambda1}).

Consider the 2-weight vector $\bv=(1,1,0,\ldots,0)$ and let $\bx\in C_4$ be a codeword covering $\bv$. Without loss of generality, assume $\bx=(1,1,x,x,0,\ldots,0)$, where $x\in\F_4^*$ (note that the coordinates of $\bx$ not covering $\bv$ must be equal because $\bx\cdot\bx=0$).

If $x\neq 1$, consider the vector $\bu=(1,x,0,\ldots,0)$ and let $\by\in C_4$ be a codeword covering $\bu$. Note that $\eta=|\supp(\bx)\cap\supp(\by)|=3$. Indeed, if $\eta=2$, then $\bx\cdot\by=1+x\neq 0$, and if $\eta=4$, then $\wt(\bx-\by)<4$. Without loss of generality, assume that $\by=(1,x,y,0,z,0,\ldots,0)$, where $y,z\in\F_4^*$. Now, we have that $\bx\cdot\by=1+x+xy$. By self-duality, $\bx\cdot\by=0$, implying $xy=x^2$ (recall that in $\F_4$, $1+\alpha+\alpha^2=0$ for $\alpha\in\F_4\setminus\{0,1\}$), and hence $y=x$. But now, $\bx+\by=(0,1+x,0,x,z,0,\ldots,0)$ which has weight less than 4 getting a contradiction.

If $x=1$, then consider the vector $\bu=(1,\alpha,0,\ldots,0)$ ($\alpha\in\F_4\setminus\{0,1\}$) and let $\by\in C_4$ be a codeword covering $\bu$. As before, $|\supp(\bx)\cap\supp(\by)|=3$ and we can assume $\by=(1,\alpha,y,0,z,0,\ldots,0)$, where $y,z\in\F_4^*$. In this case, we obtain $\bx\cdot\by=1+\alpha+y$, and since $\bx\cdot\by=0$, $y=\alpha^2$. However, $\by\cdot\by=0$ implies $1+\alpha^2+\alpha+z^2=0$, which gives $z=0$, again getting a contradiction.
\qed

The following proposition and corollary shows the nonexistence of self-dual CR codes for $k=3$ and $d=4$.

\begin{prop}\label{Noq7}
For $k\geq 3$, there is no self-dual CR $[2k,k,4]_7$ code.
\end{prop}
\pr
Consider $\F_7$ as $\Z_7$ and note that $x^2\in\{1,2,4\}$ for any element $x\in\Z_7^*=\Z_7\setminus\{0\}$. If $C$ is a self-dual CR $[2k,k,4]_7$ code, then $C_4$ is a $7$-ary $2-(2k,4,\lambda)$-design, by Lemma \ref{disseny} (where $\lambda=1$, by Lemma \ref{lambda1}).

 Consider the 2-weight vectors $\bv=(1,1,0,\ldots,0)$ and $\bu=(1,2,0\ldots,0)$ and let $\bx,\by\in C_4$ be codewords covering $\bv$ and $\bu$, respectively. Note that $\eta=|\supp(\bx)\cap\supp(\by)|=3$. Indeed, if $\eta=2$, then $\bx\cdot\by=3\neq 0$, and if $\eta=4$, then $\wt(\bx-\by)<4$. Thus, without loss of generality, assume $\bx=(1,1,a,b,0,\ldots,0)$ and $\by=(1,2,c,0,d,0,\ldots,0)$, where $a,b,c,d\in\Z_7^*$. By self-duality, on the one hand, $\bx\cdot\bx=0$, implying $a^2+b^2=5$ and hence $a^2\in\{1,4\}$. The possible values of $a$ are then $a\in\{1,2,5,6\}$. On the other hand, $\by\cdot\by=0$ implies $c^2+d^2=2$. So, $c^2=1$ and the possible values of $c$ are $c\in\{1,6\}$. Therefore, we have $ac\in\{1,2,5,6\}$.

 Finally, we obtain a contradiction taking into account that $\bx\cdot\by=0$. Indeed, $\bx\cdot\by=1+2+ac$ implies $ac=4$.
\qed

\begin{corollary}\label{Nok3}
There is no self-dual CR $[6,3,4]_q$ code with covering radius $\rho=3$.
\end{corollary}
\pr
Assume that $C$ is a self-dual CR $[6,3,4]_q$ code. By Lemma \ref{disseny}, the codewords in $C_4$ form a $q$-ary $2-(6,4,\lambda)$-design. Hence, according to Eq. (\ref{idesign}), we have:
\begin{equation}\label{A4}
  A_4=\lambda_0 = \lambda \frac{\binom{6}{2}}{\binom{4}{2}}(q-1)^2=\lambda\frac{5}{2}(q-1)^2.
\end{equation}
In this case the nonzero weights are $w_1=4, w_2=5, w_3=6$. As can be seen in the fifth row of Table \ref{taula}, $B_4=15$. Hence, $A_4=B_4 (q-1)=15(q-1)$. Comparing with Eq. (\ref{A4}) and by Lemma \ref{lambda1}, we conclude that $q=7$ and $\lambda=1$; or $q=2$ and $\lambda=6$. But this last binary case is not possible since $\lambda$ cannot be greater than $2$, by Corollary \ref{PP3}(iii).

Now, the result follows from Proposition \ref{Noq7}.
\qed

\begin{prop}\label{NoNobinarid4}
  There is no self-dual CR $[2k,k,4]_q$ code with covering radius $\rho=3$ and $q>2$.
\end{prop}
\pr
By Corollary \ref{Nok3} and Lemma \ref{lambda1},
we only have to consider the cases where $k\geq 4$ and $\lambda=1$ (recall that $\lambda=p_{2,2}-1$ by Corollary \ref{PP3}(iii)). Define the function
$$
h(q,k,\lambda')=\frac{1}{3}\cdot\frac{k(2k-1)(q-1)^2[2(2k-2)(q-1)-4-6(q-2)-3\lambda']}
{2(q^k-1)-k(q-1)[4+(2k-1)(q-1)]},
$$
which is $\beta_3^{-1}$ for $\lambda=1$, according to Corollary \ref{PP3}(iii). Therefore, $h(q,k,\lambda')$ must be a positive integer number. It can be checked that for $k>10$, the value of $h(q,k,\lambda')$ is less than 1. For $4\leq k\leq 10$, the value of $h(q,k,\lambda')$ is greater that 1 for $q\leq 25$. Hence, we have to consider $h(q,k,\lambda')$ for $4\leq k\leq 10$ and $4\leq q\leq 25$ (by Lemma \ref{q23}, $q \neq 3$). For $q,k\geq 4$, the denominator of $h(q,k,\lambda')$ is positive. Thus, in order to get the numerator positive, we need $\lambda' < [4(k-1)(q-1)-6q+8]/3$.  Computationally, we have found that the only integer values of $h(q,k,\lambda')$ for these cases are $h(4,4,0)=8$, $h(4,4,5)=2$, $h(4,4,9)=1$, $h(4,6,2)=1$ and $h(7,4,9)$.
These values would correspond to codes with parameters:
$$
  [8,4,4]_4 \mbox{ with } \lambda'=0;\;\;\;[8,4,4]_4 \mbox{ with } \lambda'=5;\;\;\;[8,4,4]_4 \mbox{ with } \lambda'=9;
$$
$$
  [12,6,4]_4 \mbox{ with } \lambda'=2;\;\;\;[8,4,4]_7 \mbox{ with } \lambda'=9.
$$

By Proposition \ref{NoQuaternary}, the codes with parameters $[8,4,4]_4$ and $[12,6,4]_4$ cannot be self-dual and CR. Finally, by Proposition \ref{Noq7}, a self-dual CR $[8,4,4]_7$ does not exist.
\qed

\begin{corollary}\label{Nod4}
  There is no self-dual CR $[2k,k,4]_q$ code with covering radius $\rho=3$.
\end{corollary}
\pr
By Proposition \ref{NoNobinarid4}, we only have to consider the binary case.
For $q=2$, the expression of $\beta_3$ in Corollary \ref{PP3}(iii) becomes:
  $$
  \beta_3=\frac{3}{2}\cdot\frac{(\lambda+1)(2^k-1)-k[2(\lambda+1)+(2k-1)]}
  {k(2k-1)(\lambda+1)[(k-1)-\lambda]},
  $$
since $\lambda'=0$ due to the fact that all weights must be even (see Lemma \ref{q23} and thus $C_5=\emptyset$).
Define the function
$$
\ell(k,\lambda)=\frac{2}{3}\cdot\frac{k(2k-1)(\lambda+1)[(k-1)-\lambda]}
{(\lambda+1)(2^k-1)-k[2(\lambda+1)+(2k-1)]},
$$
Clearly, $\ell(k,\lambda)=\beta_3^{-1}$ and must be a natural number. For $k>10$, the value of $\ell(k,\lambda)$ is less than $1$. Checking all the values for $3\leq k\leq 10$ and $0\leq\lambda\leq k-1$ (according to Corollary \ref{PP3}(iii)), the result is that only $\ell(5,2)=10$ is a natural number. It corresponds to a $[10,5,4]_2$ code. But such code cannot be self-dual, as can be seen in \cite[Exercise 9.4.2]{HP}.

The final conclusion is that there is no binary self-dual CR code with minimum distance $d=4$ and covering radius $\rho=3$.
\qed

\subsection{The case $d=3$}

\begin{prop}\label{intersect2}
If $C$ is a self-dual CR $3$-weight $[2k,k,3]_q$ code, then $|\supp(\bx)\cap\supp(\by)|= 2$, for some $\bx,\by\in C_3$.
\end{prop}

\pr
Otherwise, we would have $|\supp(\bx)\cap\supp(\by)|\in \{0,3\}$ by Lemma \ref{intersect1}. Hence, by Lemma \ref{support}, the length $2k$ should be divisible by 3 and, in fact, $2k=6$ ($9$ is odd and for $2k\geq 12$, $C$ would have more than $3$ weights). Therefore, by Proposition \ref{noexis}, the weights of $C$ would be $w_1=3$, $w_2\in\{4,5\}$, $w_3=6$. Since $k=3$, we can apply again Eqs. (\ref{PPMB1}), (\ref{PPMB2}) and (\ref{PPMB3}). 

For $w_2=4$, as can be seen in row 6 of Table \ref{taula}, $B_3=-2(q-4)$ which implies $q<4$, leading to a contradiction by Lemma \ref{q23}.

For $w_2=5$, as can be seen in the last row of Table \ref{taula}, $B_3=5$.
If $|\supp(\bx)\cap\supp(\by)|= 3$, for $\bx,\by\in C_3$, then $\bx$ is a multiple of $\by$ (otherwise, taking appropriate multiples, we would get $0<\w(\bx-\by)< 3$). Hence, $B_3 = 2$ which contradicts the result of the system.
\qed

\begin{corollary}\label{keven}
If $C$ is a self-dual CR $[2k,k,3]_q$ code with covering radius $\rho=3$, then $k=6$.
\end{corollary}

\pr
By Proposition \ref{intersect2}, there exist codewords $\bx,\by\in C_3$, such that $|\supp(\bx)\cap\supp(\by)|= 2$ and thus $|\supp(\bx)\cup\supp(\by)|= 4$. Now, if $\bz\in C_3$ has $\supp(\bz)\cap(\supp(\bx)\cup\supp(\by))\neq\emptyset$, we claim that $\supp(\bz)\subset(\supp(\bx)\cup\supp(\by))$. Otherwise, without loss of generality, assume that $\bx=(1,x_2,x_3,0,\ldots,0)$ and $\by=(1,y_2,0,y_3,0,\ldots,0)$. By Lemma \ref{intersect1}, we can assume that $\bz=(z_1,z_2,0,0,z_3,0,\ldots,0)$. Now, since $q>2$, we can take a multiple of $\bz$, say $\bz'=(z'_1,z'_2,0,0,z'_3,0,\ldots,0)$, such that $z'_2=x_2-y_2$. Hence, $\wt(\bx-\by-\bz')=4$. But we can take another multiple, say $\bz''$, such that $z''_2\neq x_2-y_2$. In this case, $\wt(\bx-\by-\bz'')=5$. So, $C$ has weights $3$, $4$ and $5$, which is a contradiction, by Proposition \ref{noexis}.

As a consequence, we have that $C_3$ induces a partition of the set of coordinates in $4$-subsets, say $P_1,\ldots, P_{k/2}$, implying that $2k$ is a multiple of $4$. Therefore, $k$ is even.

On the other hand, if we take a 2-weight vector $\bx$ with one nonzero coordinate in $P_i$ and the other one in $P_j$ ($i\neq j$), then $d(\bx,C)=2$ and $B_{\bx,2}=1$, since the zero codeword is the only one at distance $2$ from $\bx$. Now, take any two-weight vector $\by$ with both nonzero coordinates in $P_i$. Since $|P_i|=4$, there exist some codeword $\bz$ of weight $3$ including the support of $\by$ and (taking the appropriate multiple) such that $d(\bz,\by) \leq 2$. If $d(\bz,\by) = 2$, then $B_{\by,2}>1$ and the code would not be CR.  This means that any 2-weight vector $\by$ with both nonzero coordinates in $P_i$ is at distance $1$ from $C$. In other words, the projection of $C$ in $P_i$, for any $i=1,\ldots,k/2$, must be a Hamming code of length $n=(q^m-1)/(q-1)=4$, i.e. a ternary Hamming $[4,2,3]_3$ code which is self-dual (see Lemma \ref{HammingSelf}). Since $C$ has covering radius $\rho=3$, there exists some 3-weight vector $\bx$ such that $d(\bx,C)=3$. Thus, $\bx$ has the three nonzero coordinates in different $P_i$'s. This implies $k\geq 6$, but for $k>6$, $C$ would have more than three nonzero weights. As a conclusion $k=6$.
\qed

\subsection{The full classification}

Now, from Propositions \ref{D2}, \ref{4or8}, \ref{Case4}, \ref{Case6}, and Corollaries \ref{Nod5}, \ref{Nod4}, \ref{keven}, we obtain the main classification theorem.

\begin{theo}\label{main}
Let $C$ be a self-dual CR $[n,k,d]_q$ code.
\begin{itemize}
\item[(i)] If $d\leq 2$, then $C$ is the direct sum of $j$ copies ($j=1,2,\ldots$) of a $[2,1,2]_q$ code with generator matrix $(1\;\;\alpha)$ such that $\alpha^2=-1$. Such $q$-ary code exists if and only if $-1$ is a square in $\F_q$. The code $C$ has covering radius $\rho=j$ and intersection array
    $$
    \IA=\{2j(q-1),2(j-1)(q-1),\ldots,2(q-1);2,4,\ldots,2j\}.
    $$
\item[(ii)] If $d=3$ and $\rho=1$, then $C$ is the ternary Hamming $[4,2,3]_3$ code with intersection array
    $$
    \IA=\{8;1\}.
    $$
\item[(iii)] If $d=3$ and $\rho=2$, then $C$ is
\begin{itemize}
\item[(iii.i)] the direct sum of two ternary Hamming $[4,2,3]_3$ codes, that is, $C$ is a $[8,4,3]_3$ code with weights $w_1=3$ and $w_2=6$, and intersection array
$$
\IA=\{16,8;1,2\};
$$
or
\item[(iii.ii)] a $[4,2,3]_q$ code, where $q=2^r$ with $r>1$. Such code can have generator matrix
$$
G~=~\left(
\begin{array}{ccll}
~1~&~1~&~1~&~1\\
~0~&~1~&~\alpha~&~\beta\\
\end{array}
\right),
$$
where $\alpha, \beta \in \F^*_q$ are two
different elements such that $\alpha~+~\beta~+~1~=~0$. $C$ has weights $w_1=3$ and $w_2=4$ (so, an antipodal code), and intersection array
$$
\IA = \{4(q-1), 3(q-3); 1, 12\}.
$$
\end{itemize}
\item[(iv)] If $d=3$ and $\rho=3$, then $C$ is the direct sum of three ternary Hamming $[4,2,3]_3$ codes, that is, a $[12,6,3]_3$ code with weights $w_1=3$, $w_2=6$, $w_3=9$  and intersection array
$$
\IA=\{24,16,8;1,2,3\}.
$$
\item[(v)] If $d=4$ and $\rho\leq 3$, then $C$ is the extended binary Hamming $[8,4,4]_2$ code, with weights $w_1=4$ and $w_2=8$ (so, an antipodal code), and with intersection array
$$
\IA=\{8,7;1,4\}.
$$
\item[(vi)] If $d=5$ and $\rho\leq 3$, $C$ does not exist.
\item[(vii)] If $d=6$ and $\rho\leq 3$, then $C$ is the extended ternary Golay $[12,6,6]_3$ code, with weights $w_1=6$, $w_2=9$, $w_3=12$ (so, an antipodal code), and with intersection array
$$
\IA= \{24, 22, 20; 1, 2, 12\}.
$$
\end{itemize}
No other self-dual CR codes with $\rho\leq 3$ exist.
\end{theo}

\pr
(i) Direct from Proposition \ref{D2}.

(ii) In this case, since $e=\rho=1$, $C$ is a self-dual perfect single-error-correcting code. Hence, $C$ is a self-dual Hamming code and, by Lemma \ref{HammingSelf}, $C$ is the ternary Hamming $[4,2,3]_3$ code. The intersection array is trivial and can be seen, for instance, in family (F.1) of \cite{BRZ19}.

(iii) By Proposition \ref{4or8}, $C$ has length $n=4$ or $n=8$.

(iii.i) If $n=8$, let $C$ be a self-dual CR $[8,4,3]_q$ code with covering radius $\rho=2$. By the argument in the proof of Proposition \ref{4or8}, the set of coordinates $\{1,\ldots,8\}$ is partitioned into two $4$-subsets, say $A$ and $B$, such that any codeword of weight $3$ has its support contained in $A$ or in $B$. Since $C$ must be a 2-weight code, these weights are trivially $w_1=3$ and $w_2=6$. Therefore $C$ is the direct sum $C=C_1 \oplus C_2$ of two 1-weight codes (whose nonzero codewords have weight $3$). It is clear that $C$ is self-dual if and only if $C_1$ and $C_2$ are self-dual.

On the other hand, if we take a 2-weight vector $\bx$ with one nonzero coordinate in $A$ and the other one in $B$, then $d(\bx,C)=2$ and $B_{\bx,2}=1$, since the zero codeword is the only one at distance $2$ from $\bx$. Now, take any two-weight vector $\by$ with both nonzero coordinates in $A$ (or in $B$). Since $|A|=|B|=4$, there exist some codeword $\bz$ of weight $3$ including the support of $\by$ and (taking the appropriate multiple) such that $d(\bz,\by) \leq 2$ (note that $q>2$ by Lemma \ref{q23}). If $d(\bz,\by) = 2$, then $B_{\by,2}>1$ and the code would not be CR.  This means that any 2-weight vector $\by$ with both nonzero coordinates in $A$ (or in $B$) is at distance $1$ from $C$. In other words, $C_1$ and $C_2$ must be self-dual Hamming codes. By Lemma \ref{HammingSelf}, $C_1$ and $C_2$ are ternary Hamming codes of length $4$. Therefore, $C$ is the direct sum of two ternary Hamming $[4,2,3]_3$ codes.
Indeed, the direct sum of perfect codes is a CR code (see Lemma \ref{DirectSum}). The intersection array follows from (ii) and Lemma \ref{DirectSum}.

(iii.ii) If $n=4$, then $C$ is a self-dual CR $[4,2,3]_q$ code (with nonzero weights 3 and 4). By \cite[Thm. 3.3]{BRZ10}, a parity-check matrix of an antipodal CR code with $\rho=2$, up to equivalence, is a matrix with the all-ones vector in a row. In our case, this means that the generator matrix of $C$ looks as:
$$
G~=~\left(
\begin{array}{ccll}
~1~&~1~&~1~&~1\\
~0~&~1~&~\alpha~&~\beta\\
\end{array}
\right).
$$
Therefore, $1+1+1+1=0$ and thus $q=2^r$ for $r>1$, since $q > 3$, by Lemma \ref{q23}. Hence, $1+\alpha + \beta =0$. This antipodal code corresponds to the family (F.48) in \cite{BRZ19} (see \cite{BRZ10}, where self-duality is also justified when $q=2^r$, $r>1$). The intersection array can also be seen in \cite{BRZ19}.

(iv) In this case, by Corollary \ref{keven}, $C$ must be a $[12,6,3]_q$ code. By the argument of the proof of Corollary \ref{keven} (similar to the case (iii.i)), $C$ is the direct sum of three ternary Hamming $[4,2,3]_3$ codes. The intersection array follows from (ii) and Lemma \ref{DirectSum}.

(v) By Proposition \ref{Case4} and Corollary \ref{Nod4}, we have that $C$ is a $[8,4,4]_2$ code. This is the well-known binary extended Hamming of length $8$, which is self-dual. Trivially the weights are $4$ and $n=8$. This code falls into the family (F.2) in \cite{BRZ19}, where the intersection array is also specified.

(vi) Since $d=5$, we have that $e=2$. Hence $\rho>2$, otherwise $C$ would be a perfect doubly-error-correcting code. The only such code is the ternary Golay $[11,6,5]_3$ code, which obviously is not self-dual. For $\rho=3$ the code $C$ cannot exist by Corollary \ref{Nod5}.

(vii) For $d=6$, again $\rho>2$, and for $\rho=3$ we have that $C$ is a $[12,6,6]_3$ code, by Proposition \ref{Case6}. As can be seen in \cite{DG73,Pless68}, any code with these parameters must be the extended ternary Golay code, which is self-dual. The weights of such code are $6$, $9$ and $12$, as can be seen, for example, in \cite{MS}. This code corresponds to (S.12) in \cite{BRZ19}, where the intersection array is also specified.
\qed

\section{Concluding remarks}\label{concluding}

Let $q'$ be a prime power such that $-1$ is a square in $\F_{q'}$, and let $q''=2^r$ with $r>1$. Then,
from Theorem \ref{main}, we see that the parameters for self-dual CR codes are
\begin{itemize}
  \item For $\rho=1$: $[2,1,2]_{q'}$, $[4,2,3]_3$.
  \item For $\rho=2$: $[4,2,2]_{q'}$, $[4,2,3]_{q''}$, $[8,4,3]_3$, $[8,4,4]_2$.
  \item For $\rho=3$: $[6,3,2]_{q'}$, $[12,6,3]_3$, $[12,6,6]_3$.
\end{itemize}

For $\rho=4$, obviously we have the codes with parameters $[8,4,2]_{q'}$ and $[16,8,3]_3$, corresponding to the direct sums of four copies of a self-dual $[2,1,2]_{q'}$ code and four copies of the ternary Hamming $[4,2,3]_3$ code. In addition, we have the binary extended Golay $[24,12,8]_2$ code. For $\rho > 4$, apart from the direct sums of copies of a self-dual CR code with $\rho=1$, it seems that there are no other possibilities. However, the techniques used here become of high complexity for $\rho> 3$.

\section*{Acknowledgmements}
This work has been partially supported by the Spanish Ministerio de Ciencia e Innovación  under Grants PID2022-137924NB-I00
(AEI/FEDER UE), RED2022-134306-T, and also by the Catalan AGAUR Grant 2021SGR-00643.
The research of
the second author of the paper was carried out at the
IITP RAS within the program of fundamental research on the topic
"Mathematical Foundations of the Theory of Error-Correcting Codes"
and was also supported by the National Science Foundation of
Bulgaria under project no. 20-51-18002.


\begin{thebibliography}{99}

\bibitem{BZZ77}
L.A. Bassalygo, V.A. Zinoviev,
``A note on uniformly packed codes",
{\em Problems Inform. Transmiss.}, vol. 13, no. 3, pp. 22--25, 1977.

\bibitem{BZZ74}
L.A. Bassalygo, G.V. Zaitsev, V.A. Zinoviev, ``Uniformly packed
codes," {\em Problems Inform. Transmiss.,}  vol. 10, no. 1, pp.
9--14, 1974.

\bibitem{BRZ10} J. Borges, J. Rifà, V. A. Zinoviev,
``On $q$-ary linear completely regular codes with $\rho=2$
and antipodal dual", {\em Adv. in Maths. of Comms}.,
vol. 4, pp. 567--578, 2010.

\bibitem{BRZ19} J. Borges, J. Rif{\`{a}}, V.A. Zinoviev,
``On completely regular codes", {\it Problems Inform. Transmiss.}, vol. 55, no. 1, pp. 1--45, 2019.

\bibitem{rho2} J. Borges, V.A. Zinoviev, D.V. Zinoviev,
``On the classification of completely regular codes
with covering radius two and antipodal duals", {\it Problems Inform. Transmiss.}, vol. 59, no. 3, pp. 204--216, 2023.

\bibitem{BCN89}
A.E. Brouwer, A.M. Cohen and A. Neumaier, {\em Distance-Regular
Graphs}, Springer, Berlin, 1989.

\bibitem{DKT}
E.~R. van Dam, J.H. Koolen, H. Tanaka, ``Distance-Regular graphs"
    {\em The Electronic Journal of Combinatorics.}, DS. no. 22, pp. 1--156.

\bibitem{D73} P. Delsarte, ``An Algebraic Approach to the
Association Schemes in Coding Theory", {\em Philips Res. Rep. Suppl.}, 10, 1973.

\bibitem{DG73} P. Delsarte, J.M. Goethals, ``Unrestricted codes with the Golay parameters are unique,"
{\em Discrete Math.}, vol. 12, pp. 211–-224, 1975.

\bibitem{GvT} J.M. Goethals, H.C.A. Van Tilborg, ``Uniformly packed codes", {\em Philips Res.}, vol. 30, pp. 9--36, 1975.

\bibitem{HP} W.C. Huffman, V. Pless, {\em Fundamentals of error-correcting codes}, Cambridge Univ. Press, New York, 2003.

\bibitem{KKM}
J.H. Koolen, D. Krotov, B. Martin, {\em Completely regular codes}, 2016,
    \emph{https://sites.google.com/site/completelyregularcodes}.

\bibitem{MS} F.J. MacWilliams, N.J.A. Sloane, {\em The theory of error-correcting codes}, North Holland Publishing Company, 1977.

\bibitem{N92}
A. Neumaier, ``Completely regular codes," {\em
Discrete Maths.}, vol. 106/107, pp. 335---360, 1992.

\bibitem{Pless} V. Pless, ``Power moment identities on weight distributions in error correcting codes,"
{\em Inform. and Control}, vol. 6, pp. 147-–152, 1963.

\bibitem{Pless68} V. Pless, ``On the uniqueness of the Golay codes," {\em J. Comb. Theory}, vol.5, pp. 215-–228, 1968.

\bibitem{SZZ71} N.V. Semakov, V.A. Zinoviev, G.V. Zaitzev, ``Uniformly packed
codes", {\em Problems Inform. Transmiss.}, vol. 7, pp. 38--50, 1971.

\bibitem{S90} P.~Sol{\'e}, ``Completely regular codes and
completely transitive codes",
{\em Discrete Maths.}, vol.81, no. 2 (1990), pp. 193--201, 1990.

\bibitem{Tiet} A. Tiet\"av\"ainen, ``On the Nonexistence of Perfect Codes over Finite Fields", {\em SIAM J. Appl. Math.}, 1973,
vol. 24, no. 1, pp. 88-–96, 1993.

\bibitem{ZL73} V.A. Zinoviev, V.K. Leontiev, ``On Non-existence of Perfect Codes over Galois Fields", {\em Probl.
Upravlen. Teor. Inform.}, 1973, vol. 2, no. 2, pp. 123–-132 [Probl. Control Inf. Theory (Engl. Transl.),
1973, vol. 2, no. 2, pp. 16-–24].

\end{thebibliography}
\end{document}